# Exploration of terahertz from time-resolved ultrafast spectroscopy in single-crystal $Bi_2Se_3$ topological insulator

Prince Sharma[1, 2], Mahesh Kumar[1, 2], V.P.S. Awana[1, 2*]

[1]*CSIR-National Physical Laboratory, Dr. K.S. Krishnan Marg, New Delhi-110012, India*
[2]*Academy of Scientific and Innovative Research (AcSIR), Ghaziabad, 201002, India*

**Abstract**

In this article, we reconnoiter the differential reflection signal of a $Bi_2Se_3$ single crystal flake, using ultrafast transient absorption spectroscopy in the femtosecond time domain and thereby explore the experimental data in terms of terahertz frequency generated in the sample. An exfoliated flake of a well characterized self-flux grown bulk $Bi_2Se_3$ single crystal having rhombohedral structure and layered morphology is used in the present study. The kinetic profile of the same being generated through a reflection signal by a pump laser of 650 nm at an average power of 0.5 mW is studied utilizing time-resolved ultrafast technique. The silhouette as a function of probe delay predicting the capability of the terahertz generation is estimated. Here, two methods FFT (fast Fourier transformation) and FFD (filtering high-frequency component followed by fitting data) are performed to estimate the value of terahertz generated in the system. While comparing the two (FFT & FFD) it is found that a large amount of magnitude difference occurs in the prediction of terahertz frequency. Summarily, we not only report the generation of terahertz in $Bi_2Se_3$ flake, also but points out that the exact order of magnitude and the capability of the same depends upon the method of analysis. It is important to extract the vibration signal from the background one so that to find the exact order of magnitude and capability of terahertz generation by any quantum material.

***Corresponding Author**

Dr. V. P. S. Awana: E-mail: awana@nplindia.org
Ph. +91-11-45609357, Fax-+91-11-45609310
Homepage: awanavps.webs.com

**INTRODUCTION**

A new quantum state of matter having conducting surface state and an insulating bulk energy band structure has attained a lot of interest of the research community[1–5].These fascinating compounds are called topological insulators (TI), of whom astonishing surface state (SS)are protected by time-reversal symmetry (TRS) and spin-orbit coupling (SOC)[1–7]. This non-trivial conducting state has a one-to-one corresponding relation between energy and momentum in a low energy region, which is termed as Dirac cone[1, 2, 8, 9]. The spin-polarized characteristics of the surface states of the TIs do have a large potential in the field of quantum computation, spintronics as well as in the field of nonlinear optics (NLO)[4, 10–12].Further, TIs had gained tremendous momentum in the field of optical nonlinearity in terms of their intriguing applications such as terahertz generation/ detection, frequency shifting, optical signal processing, optical switching, optical power limiters and optical data storage for information processing[13–23]. The exploration of the optical properties, electron-phonon interaction, carrier dynamics, low-energy electronic responses, etc., of TIs, has aggravated to exploit these materials for optimizing as high frequency and high-intensity terahertz generators and optoelectronic device performance[14, 17, 24, 25]. Ultrafast statistics of carriers and phonons in TI were studied by time-resolved ultrafast spectroscopy (TRUS). It is one of the standard tools for studying the excited state dynamics of the system.

Further, the dynamic properties of hot carriers in TI that are excited by pump laser have engrossed a great deal of attention as the three-dimensional single crystal of topological insulator generates terahertz frequency[17, 26–30].The terahertz frequency has an effervescent role in numerous applications such as bio-imaging, medical diagnoses, T-bit communications, security cameras, explosive detection and industrial inspections[26, 31–33]. Terahertz frequency is generated by the hot carriers in a single crystal of TI ($Bi_2Se_3$) as the system is pumped by the laser in its bulk conduction band. These hot carriers while relaxing cause phonon oscillations in optical as well as the acoustic regime. The order of terahertz frequency can be experimentally calculated using the differential reflection signal that is generated in TRUS. The differential reflection signal as a function of probe delay is then theoretically analyzed by various approaches. A detailed study of terahertz generation of single-crystal $Bi_2Se_3$ is taken up and its theoretical interpretation by two different fitting models of experimental data are reported in this article. First, by FFT (fast Fourier transformation) and another by FFD (filtering high frequency component followed by fitting data)[14, 17, 24, 25]. The two distinct theoretical approaches (FFT & FFD) resulted in elucidation of terahertz frequency differing by an order of magnitude which has also been reported for perovskite, such as, two-dimensional lead-free hybrid halide perovskite using super atom anions with tunable electronic properties; stable dynamics performance and high efficiency of ABX3-type super-alkali perovskites first obtained by introducing $H_5O_2$ cation; tunable electronic structures and high efficiency obtained by introducing super alkali and super halogen into AMX3-type perovskites and in $Bi_2Te_3$ topological insulator[14, 17, 24–26, 34–37]. The current article not only substantiates terahertz generation from flakes of bulk single-crystal $Bi_2Se_3$, but also points out an important fact that various fitting models may lead to different results. Caution is required before using available fitting models.

## EXPERIMENTAL DETAILS

Single crystal of $Bi_2Se_3$ TI is grown by a solid-state reaction route. The purest form (>4N) of Bi (Bismuth) and Selenium (Se) were taken as precursor to the grow the crystal via the self-flux method[38, 39]. The encapsulated weighed and inert atmosphere well ground rectangular bar sample was kept in a fully automated tube furnace under a well optimized heat treatment as described in fig 1(a) [34,35]. After, this vigorous heating process, a silver-shiny 2 cm long single crystal is grown as shown in fig 1(b). In order to confirm the structural characteristics of single crystal, room-temperature X-ray diffraction (XRD) using Rigaku made Mini Flex II of Cu-$K_\alpha$ radiation ($\lambda$=1.5418 Å) was performed. SEM and EDS are performed on Bruker made scanning electron microscope. The analysis of vibration modes in $Bi_2Se_3$ crystal has been done using the Renishaw Raman spectrometer. The ultrafast transient absorption spectroscopy system was used for the measurement of differential reflection signals. This system consists of a Ti: Sapphire based femtosecond mode-locked laser (MICRA by coherent), amplifier (legend by Coherent) and spectrometer (Helios by ultrafast systems). The laser beams of 800 nm coming from amplifier are a Gaussian pulse with a spectral width of 60 nm and a repetition rate of 1 kHz. This amplified beam splits into 70:30 proportions and the large one is used as a pump beam fed to the operational parametric amplifier (OPA from light conversion). In the Helios spectrometer, a probe beam is steered through an 8 ns long delay stage and, thereafter, strikes on sapphire/$CaF_2$ crystal to generate a white light continuum. This provides a wide range of spectral profiles from 320 nm- 1600 nm.

## THEORETICAL DETAILS

The TRUS system is used for the measurement of differential reflection signals as a function of the probe beam. When $Bi_2Se_3$ single crystal is pumped by the laser, the hot carriers in bulk conduction band are excited and thereby thermalization between these hot carriers and phonons takes place. These statistics can be easily divided into two halves. One of them is thermalization between these carriers and optical phonons in a sub-picosecond timescale and the other one is in between carriers and acoustic phonons in subsequently slow component of time scale than the optical phonons[25]. The focus of this article is the generation of terahertz frequency being induced from pump laser created optical phonon vibrations in the system. The visualization of the vibration of phonons can be easily done by Fourier analysis. But the fact is that the analysis can be done by numerous approaches and different theoretical analysis gives different terahertz frequency. The two approaches that have been taken here are FFT and FFD[13, 14, 16, 17, 22, 24, 25]. In FFT, the oscillations in the sub-picosecond regime (1 ps-8 ps) are analyzed by transforming the time function to the frequency domain using Origin software version 9.1. The Origin software uses the FFTW library (Fastest Fourier Transform in the West) to compute the Fourier transformation. FFT is the fastest as well as a convenient way to compute DFT (discrete Fourier transform). A DFT converts a frequency signal to its frequency counterpart; if $x_i$ be the sequence with a length of N. The DFT function is

$$F_n = \sum_{i=0}^{N-1} x_i e^{\frac{-2\pi i}{N} n_i} \tag{1}$$

The second approach to attribute the terahertz frequency generation is to perform FFD. In FFD, the high-frequency component is extracted out from the raw data generated by TRUS followed by its damped-sinusoidal fitting. The abstraction is done by filtering out the oscillations from the curve by high-pass Fourier filter with a cut-off frequency of 2.32 THz in Origin software. The process of selecting the frequency from the signal using Fourier transforms to analyze the input is known as filtering. After filtering out the signal, the high-frequency components are fit by using damped sinusoidal function,

$$Y = Ae^{-(t-t_0)/\tau}\sin[2\pi f(t-t_1)] \quad (2)$$

Where A is the amplitude, $\tau$ is the decay time, $f$ is the frequency and Y is corresponding to ΔR/R.[25]

The fitting equation gives out the approximate frequency generated in the system. These two methods include various steps and are also described in the flow chart fig. 2(a) and 2(b). These two approaches are used for the exploration of the experimental measurements by TRUS.

## RESULT AND DISCUSSION

The XRD pattern of crystal flake on its surface and in its powder form (PXRD) is shown in fig. 3(a) and 3(b) respectively. The peaks at (00L) in the XRD pattern of crystal surface confirm the formation of the Single crystalline structure. Further, the Rietveld refinement of the PXRD pattern of $Bi_2Se_3$ is done using the FullProf Suite toolbar, confirming the rhombohedral crystal structure with R-3m (D5) space group. The Rietveld refinement gives the atom positions to be Bi (0,0,0.3999(5)), Se1(0,0,0) and Se2(0,0,0.207(4)) with lattice parameters a=b=4.14(2) $A^o$ and c=28.7(1) $A^o$. The full description of the crystal structure and its Rietveld refinement is described in our previous report [40]. Using these parameters and lattice constant, the unit cell structure of $Bi_2Se_3$ single crystal is described with the help of VESTA software and is shown in fig. 3(c). It can be seen in fig. 3(c), that this TI contains three bi-layers of Bi and Se stacked monolayers of either Bi or Se in a closely packed FCC lattice structure[38, 39]. The vibration modes of $Bi_2Se_3$ single crystal are recorded using Raman spectroscopy at room temperature. The peaks around 72.1, 131.2 and 177.1cm$^{-1}$ correspond to $A_{1g}^1$, $E_g^2$, and $A_{1g}^2$ respectively as shown in fig 3(d). These three distinct Raman active modes were in good agreement with the earlier reported results[38–41]. The experimental evidence of $Bi_2Se_3$ single crystal shows a layered crystalline structure as depicted by SEM studies. Through EDS analysis confirmed the quantitative amounts of constituent atoms in the studied single crystal close to the stoichiometric ratio and is depicted in fig 4(b).

A flake of bulk $Bi_2Se_3$ crystal being grown through self-flux method is used for excited state dynamics studies. The TRUS measurements are done using 650 nm as a pump beam (400-micron spot size and a pulse duration of 70 fs with a repetition rate of 500 Hz which give a peak power of 22.71 GW/cm$^2$) with probing in the NIR region. The time-domain data has been collected from few femtoseconds to 6 ns. The kinetic profile as shown in fig. 5(a) has been obtained at an average power of 0.5 mW as measured by power meter at point of interaction. This silhouette is being fitted using the surface Xplorer software as shown in fig. 5(b), elucidating the optical excitation of charge carriers of $Bi_2Se_3$ single crystal flake. The pump energy lies in the higher regime than the envisaged bulk electronic energy band gap of $Bi_2Se_3$. The exited electron hastily decays to its lower-lying energy state via inter and intraband phonon-mediated scattering processes, which is probed by the NIR continuum as described in kinetic profile. A metastable population of electrons occur within 1.2ps, filling the surface state with a steady supply of charge carriers by the agitated beam and then exponential decay up to 8 ps. This occupation of a metallic state is in fact a unique situation. The decay kinetic from 1.2 ps to 8 ps consists some sort of oscillations in the signal which corresponding to the optical phonons vibration of the flake. As we engrossed the kinetic profile in this smaller time scale, the oscillations are relatively faster than the other vibrations in higher time scale[13, 14, 18, 19, 22, 25]. It is clear from fig. 5(a) that there are oscillations in the kinetic profiles from 1ps to 10 ps and 10 ps to 50 ps, which happens due to the vibrations of lattice phonons. The thermalization between hot electrons and optical phonons gives upsurge to the fast component of the sub-picosecond time scale. On the other hand thermalization between acoustic phonons and electrons gives rise to a slow component of several picoseconds[14, 16, 17, 20]. The slow oscillations components between 10 ps to 50 ps correspond to the acoustic phonons vibration in the lattice and the fast oscillations between 1 to 10 ps are attributed to optical phonons as shown in fig 5(a) [24, 25, 40]. The frequency of the fast component can be assigned as the $A_{1g}^1$ COP (coherent optical phonons) mode of $Bi_2Se_3$ as prescribed in Raman spectroscopy and the slow component is attributed to the CAP (coherent acoustic phonons). The order of magnitude of terahertz in COP is higher than CAP [24]. Here, we focus on the fast components (COP mode, Fig. 6(a)) and thereby its interpretation in terms of terahertz frequency generation. The exploration of the oscillations is being done by two different methods, i.e., (a) FFT and (b) FFD. These two methods include various steps as described in the flow chart in fig. 2(a) and 2(b).

The FFT analysis of the fast oscillations component being described in fig. 6(b) shows the generation of 0.1 THz of frequency with a magnitude of 0.3 by the studied $Bi_2Se_3$ crystal. This analysis is done by following the flowchart described in figure 2(a), where the kinetic signal of the flake is cropped from 1 ps to 10 ps in order to only considering the optical phonon vibration of the flake. These induced oscillations are generated in the flake through the high pulsed laser of order of few gigawatts per $cm^2$. The cropped oscillations are then plotted in the origin software and using the FFT command in it, a complete description of the pulsations in the experimental data is taken out. It is quite clear that the highest terahertz frequency generated in the system is 0.1 THz. This much low frequency subsequently is not in accordance to the capability of the crystal as the data that is analyzed through this kind of interpretation include both the pulsations as well as exponential decay profile due to thermalization of the charge carriers. This convolution of the different effect in one data reduces the accuracy of the analysis. Thereby, a secondary analysis of the data is carried out which is termed as FFD.

The second analysis is FFD, in which the cropped data is firstly filter out from the exponential decay profile and thereafter fitted using damped sinusoidal function as described in flow chart of figure 2(b). The fast oscillations extracted data shown in fig. 6(a) is first filtered out from the decay profile. The slow varying components are filtered out from the vibrations using high-pass Fourier filter with a cut off frequency of 2.32 THz as shown in fig. 7(a) using the origin software. This flittered data is then fitted using a damped sinusoidal equation as depicted in fig. 7(b) for finding out the value of terahertz frequency generated in the flake. Fig. 7(c) shows a picture of the fitting data depicting the accuracy of the overlapping in the fitting equations and experimental data. Various parameters that come out from this fitting are described in table I. The value of frequency is easily calculated from the fitting parameters. While comparing the fitting equation as described in table I with the equation (2), it is established that frequency in terms of 'w' is given by

$$f = \frac{1}{2w}$$

After calculating the frequency from the formula, the value of terahertz frequency generated by the single crystal is 2.69±0.01THz. The terahertz frequency being generated in the crystal is analyzed using FFD, but still, it is in time-domain scale. Further, these FFD fitted data are Fourier transformed into the frequency domain to verify the terahertz frequency that is governed by the FFD analysis. Figure 8, shows the Fourier transform of the FFD fitted data and it is concluded that the crystal is capable of generating the terahertz frequency. Interestingly, this is almost equivalent to that one as determined by the time domain FFD analysis. The value of frequency after the Fourier transform of the FFD fitted data is 2.64± 0.01 THz**.** It is clear by the analysis of the TRUS kinetic profile data of the studied $Bi_2Se_3$ single crystal, that although the generation of terahertz frequency is evidenced via both models, the relative amount is different.

## CONCLUSION

We presented the TRUS of an exfoliated flake of bulk $Bi_2Se_3$ single crystal grown and studied the non-linear optical response of the same in the terahertz domain. The two different approaches i.e., FFT & FFD were used to extract the terahertz frequency from the experimental data and it is found that the former (FFT) and later (FFD) analysis results into 0.1 THz and 2.69±0.01THzrespectively at a same fluence of 0.5 mW (average power). The FFD data is obtained by using a high-frequency filter, which removes the noise along with the background incident signal, and thus results in better and improved values of the THz oscillations frequency. It is concluded that it is the necessity to remove the background signal from the output signal in order to get the precise and more accurate value of the frequency generated in TI crystals. Further, it is proposed that real time-domain terahertz spectroscopy has to be performed in order to further use this study for terahertz detector applications.

**Figure Caption**

Figure 1(a): Schematic diagram of heat treatment of grow $Bi_2Se_3$single crystal.

Figure 1(b): Silvery shine 2 cm long single crystal

Figure 2(a): Stepwise formulation of the FFT calculations

Figure 2(b): Stepwise formulation of the FFD calculations

Figure 3(a): X-ray diffraction pattern for $Bi_2Se_3$single crystals.

Figure 3(b): Rietveld fitted room-temperature X-ray diffraction pattern for powder $Bi_2Se_3$ crystals.

Figure 3(c): Unit cell structure of $Bi_2Se_3$single crystals.

Figure 3(d): Raman spectra of $Bi_2Se_3$single crystals at room temperature.

Figure 4(a): Show layered morphology of $Bi_2Se_3$ through Scanning electron microscopic.

Figure 4(b): Elemental analysis of $Bi_2Se_3$ single crystals by Energy-dispersive X-ray spectroscopy.

Figure 5(a): Differential reflection signal of the $Bi_2Se_3$ single crystals by TRUS.

Figure 5(b): Fitted kinetic profile of the $Bi_2Se_3$ single crystals by Surface Xplorer.

Figure 6(a): Cropped reflection signal as a function of probe delay.

Figure 6(b): Fast Fourier transformation (FFT) of cropped experimental data.

Figure 7(a): optical phonons oscillations by filtering the high-pass Fourier filter with a cut-off frequency of 2.32 THz.

Figure 7(b): Damped sinusoidal fitting of the optical phonons vibrations.

Figure 7(c): Skyrocketed out experimental fitting data.

Figure 8: Fourier transform analysis of the fitted damped sinusoidal equation.

**Figure 1(a)**

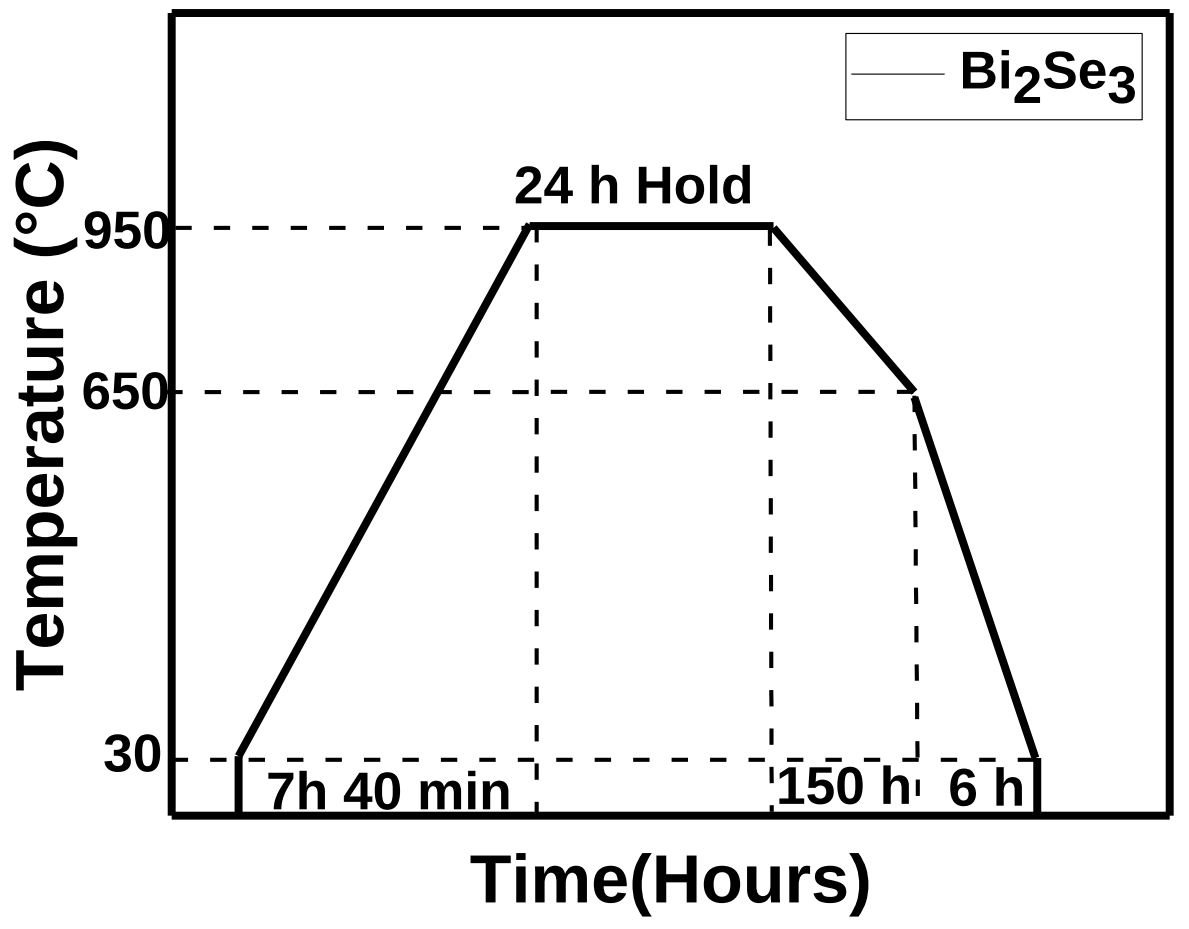

**Figure 1(b)**

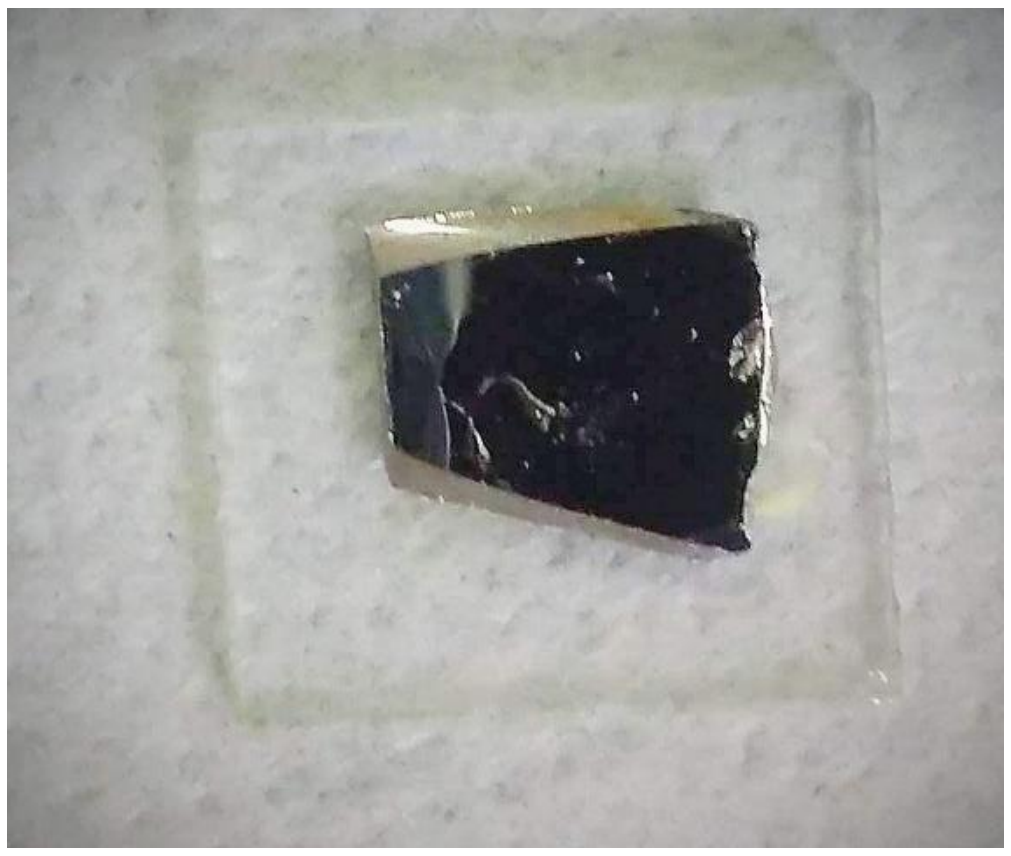

**Figure 2(a)**

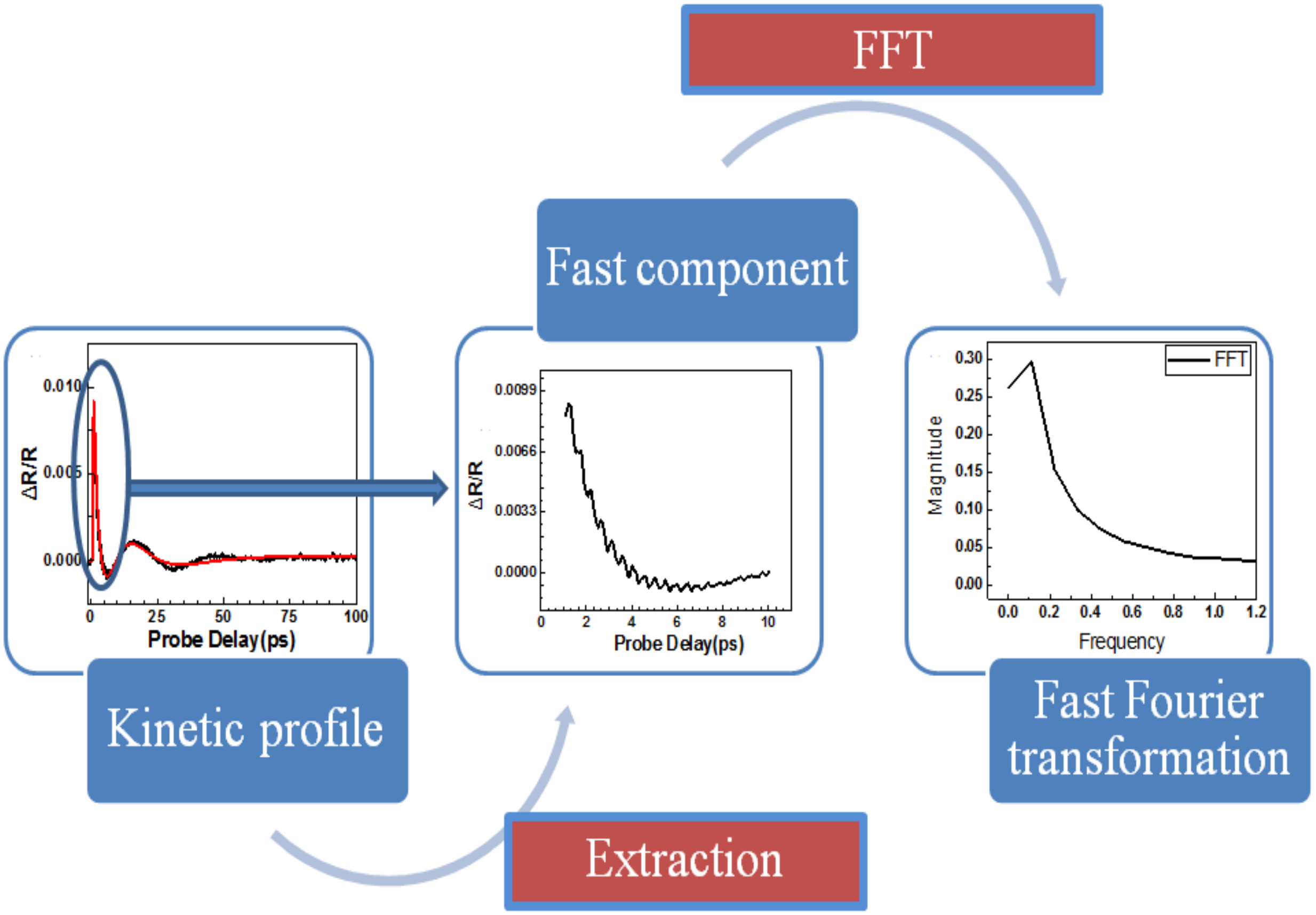

**Figure 2(b)**

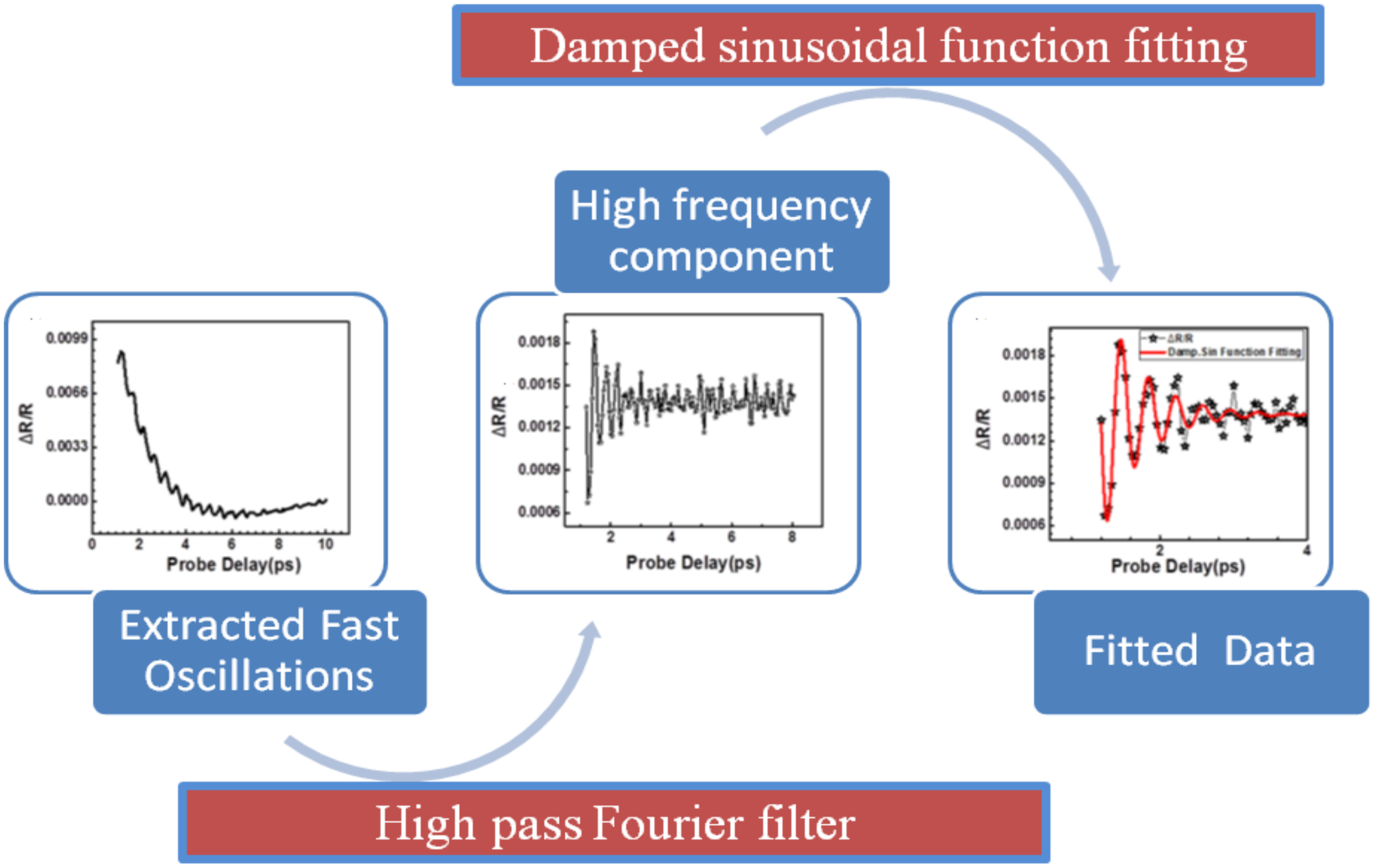

**Figure 3(a)**

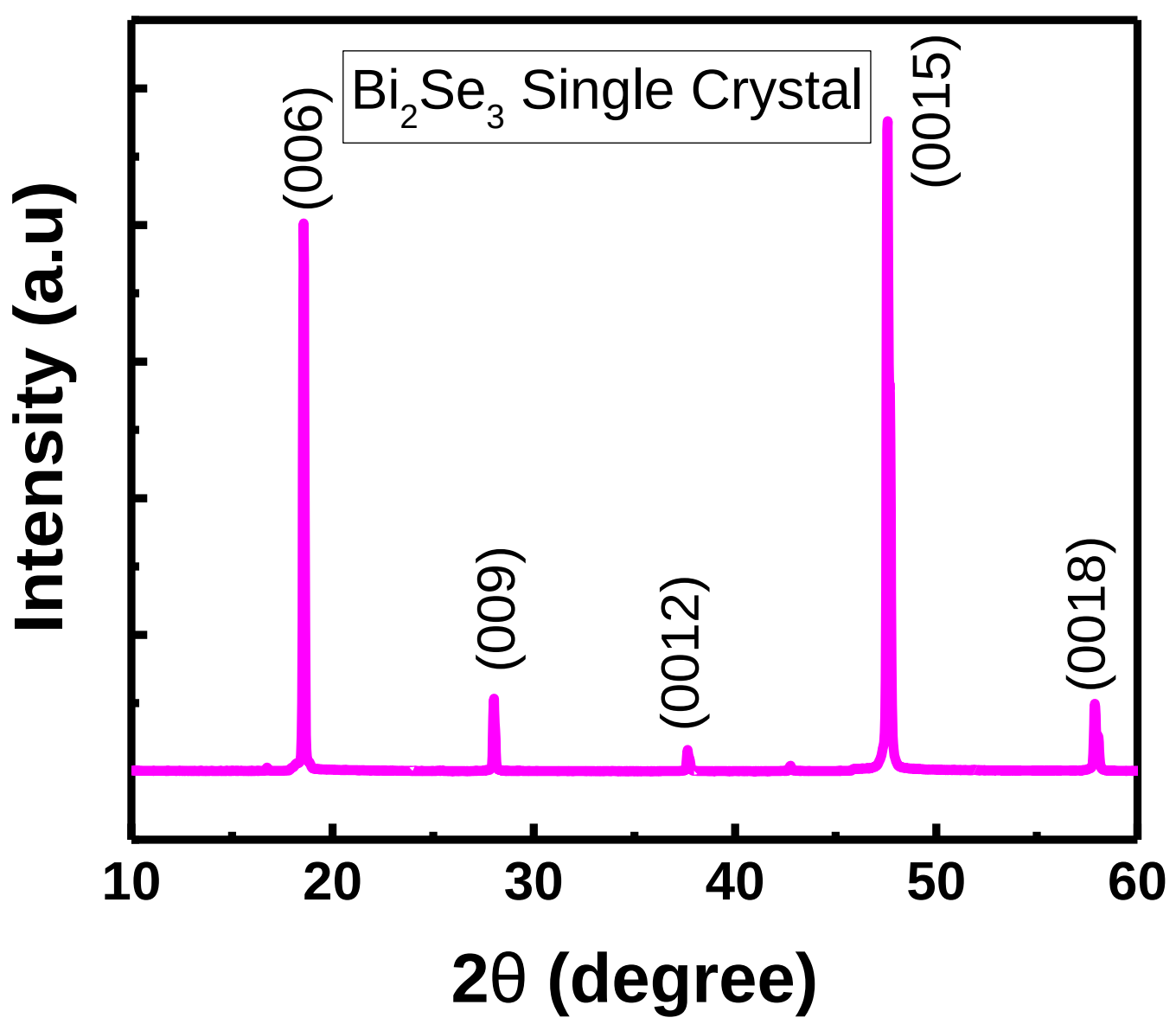

**Figure 3(b)**

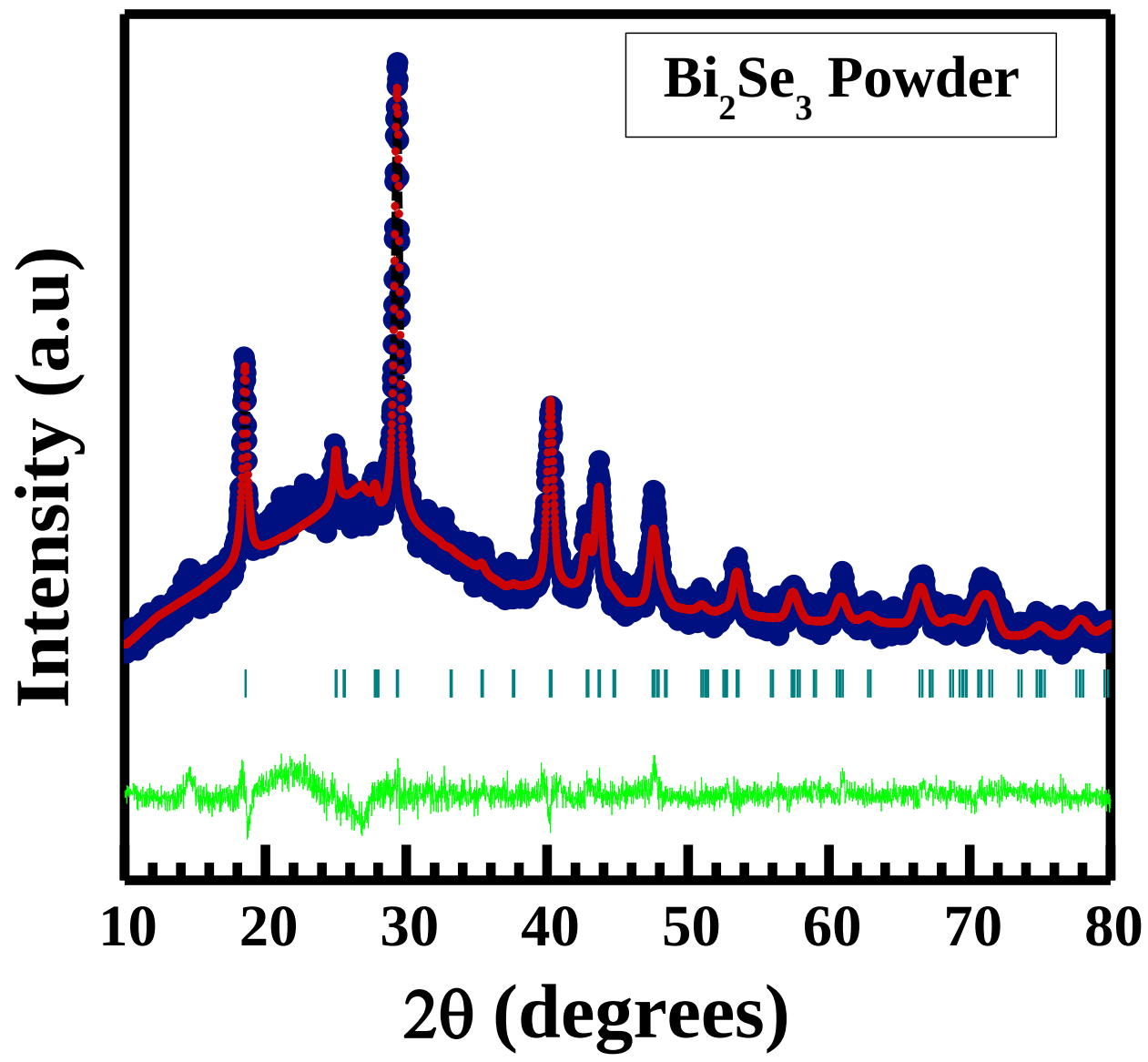

## Figure 3(c)

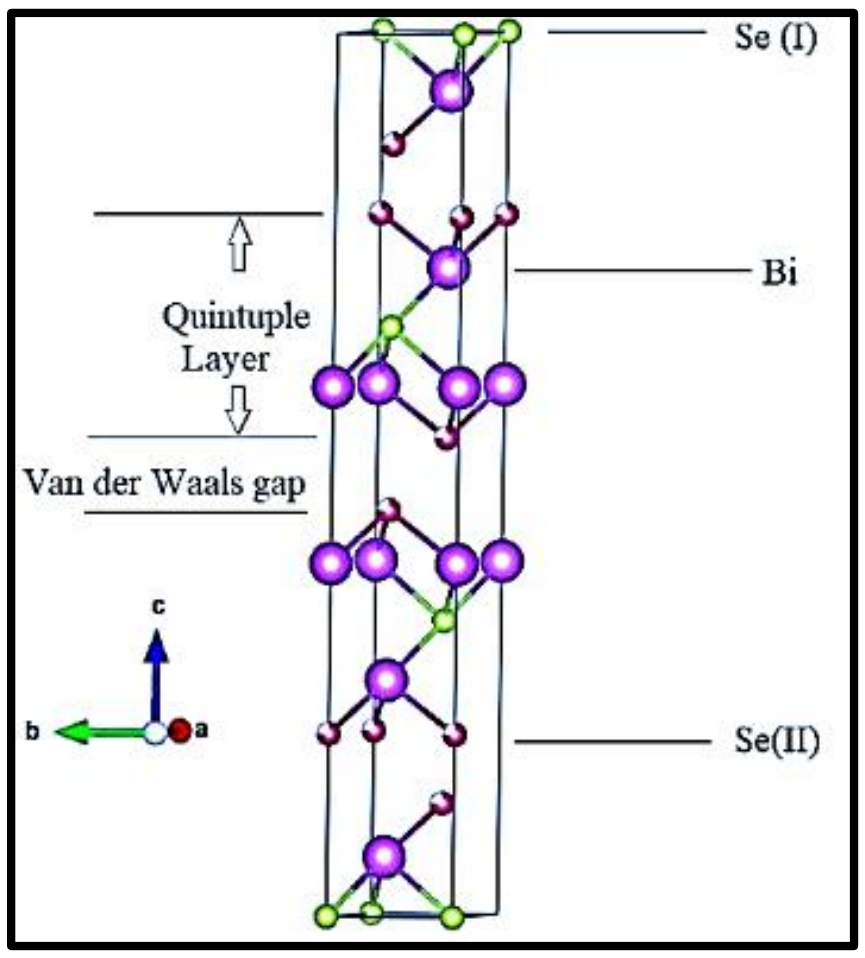

## Figure 3(d)

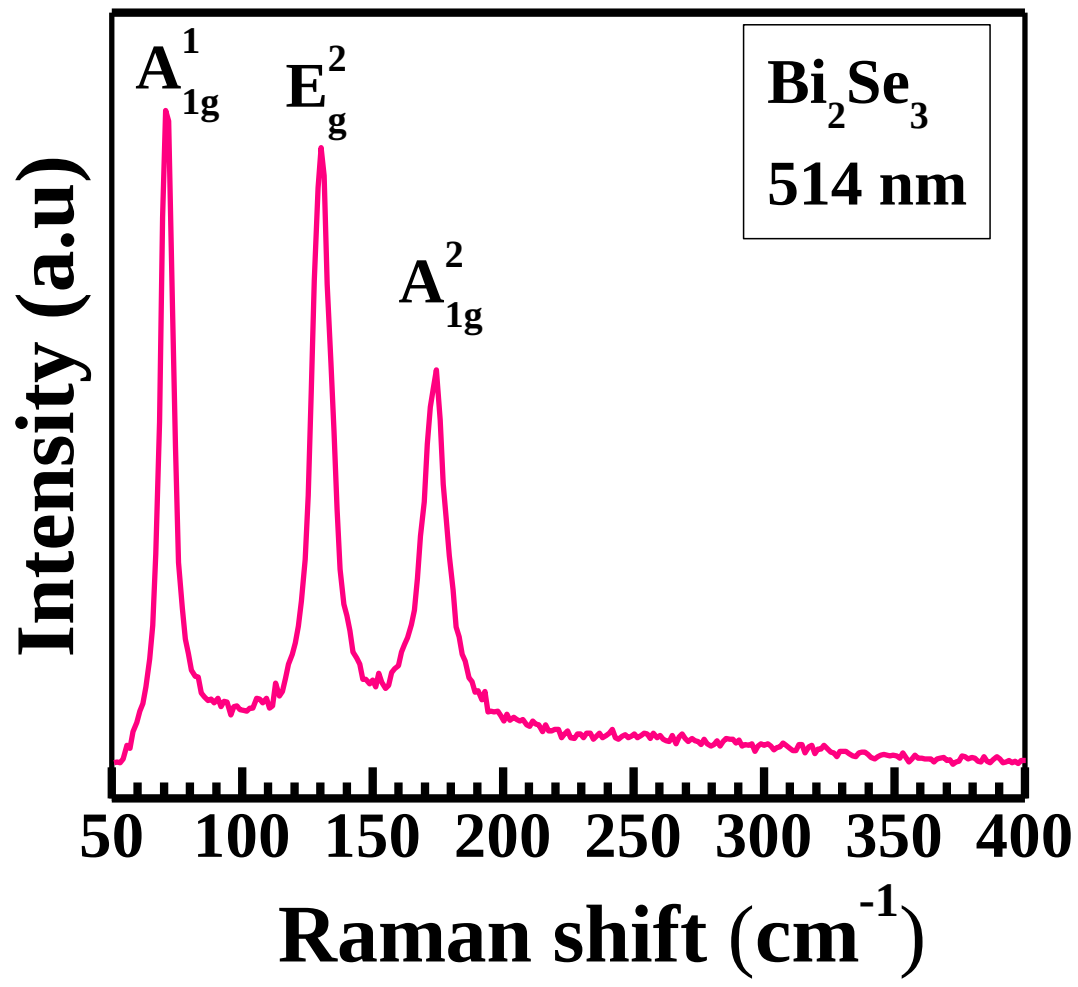

## Figure 4(a)

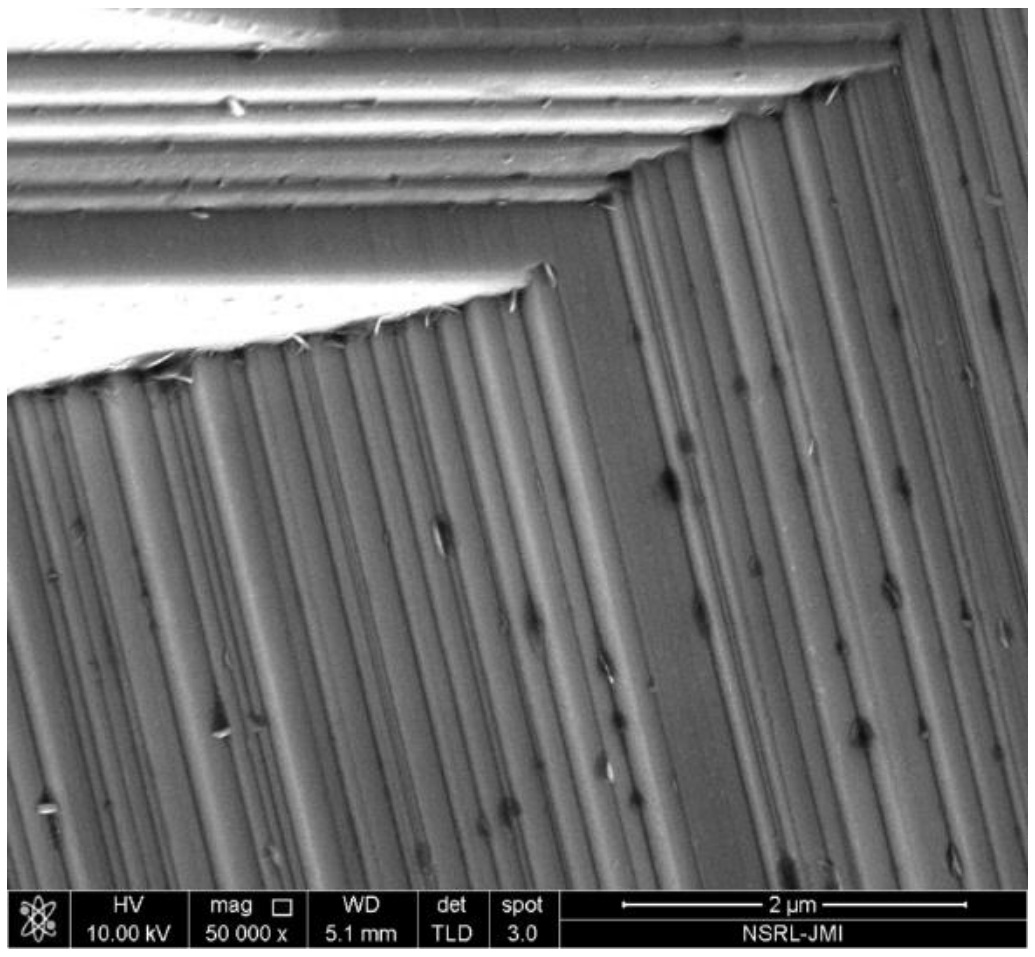

## Figure 4(b)

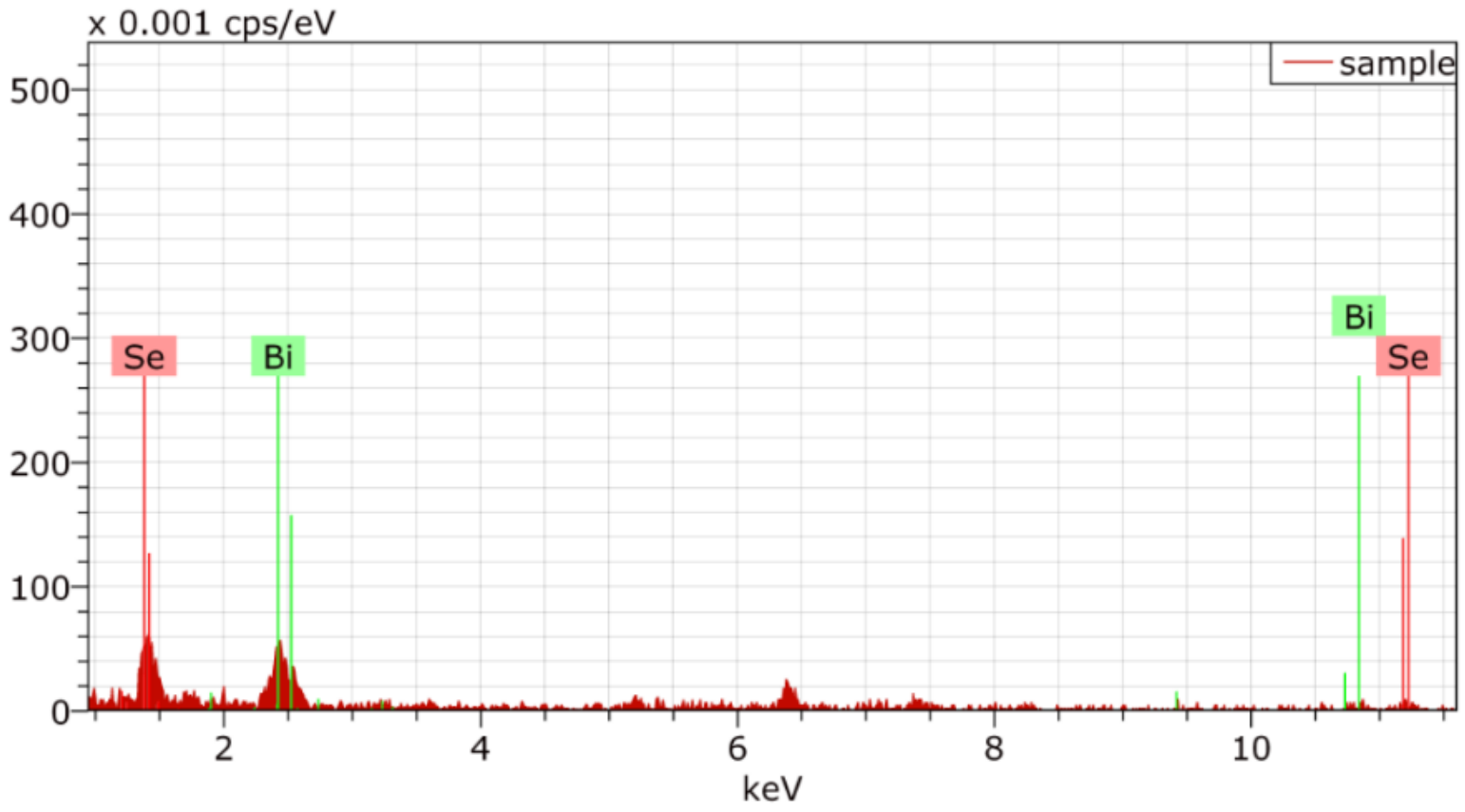

**Figure 5(a)**

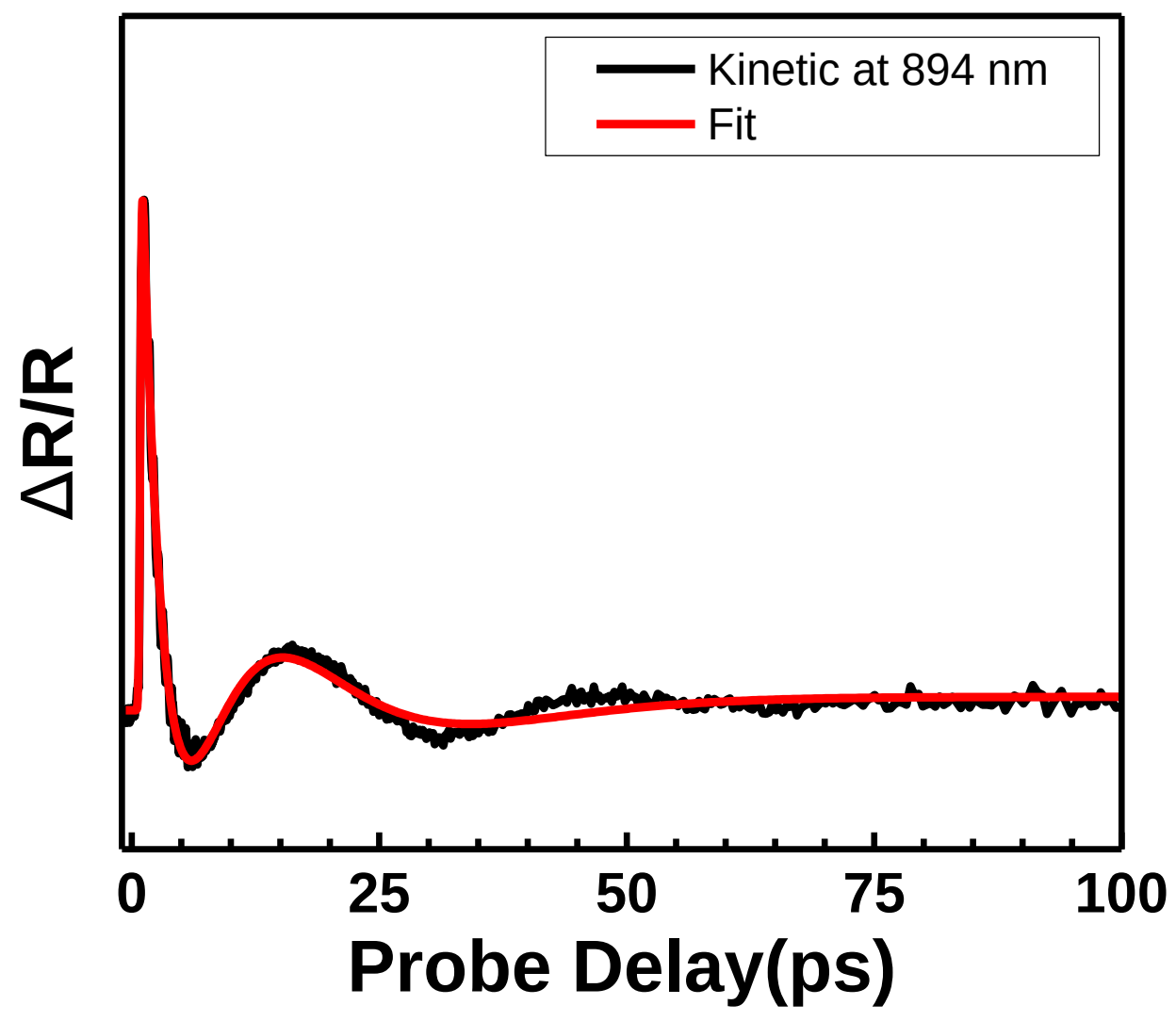

**Figure 5(b)**

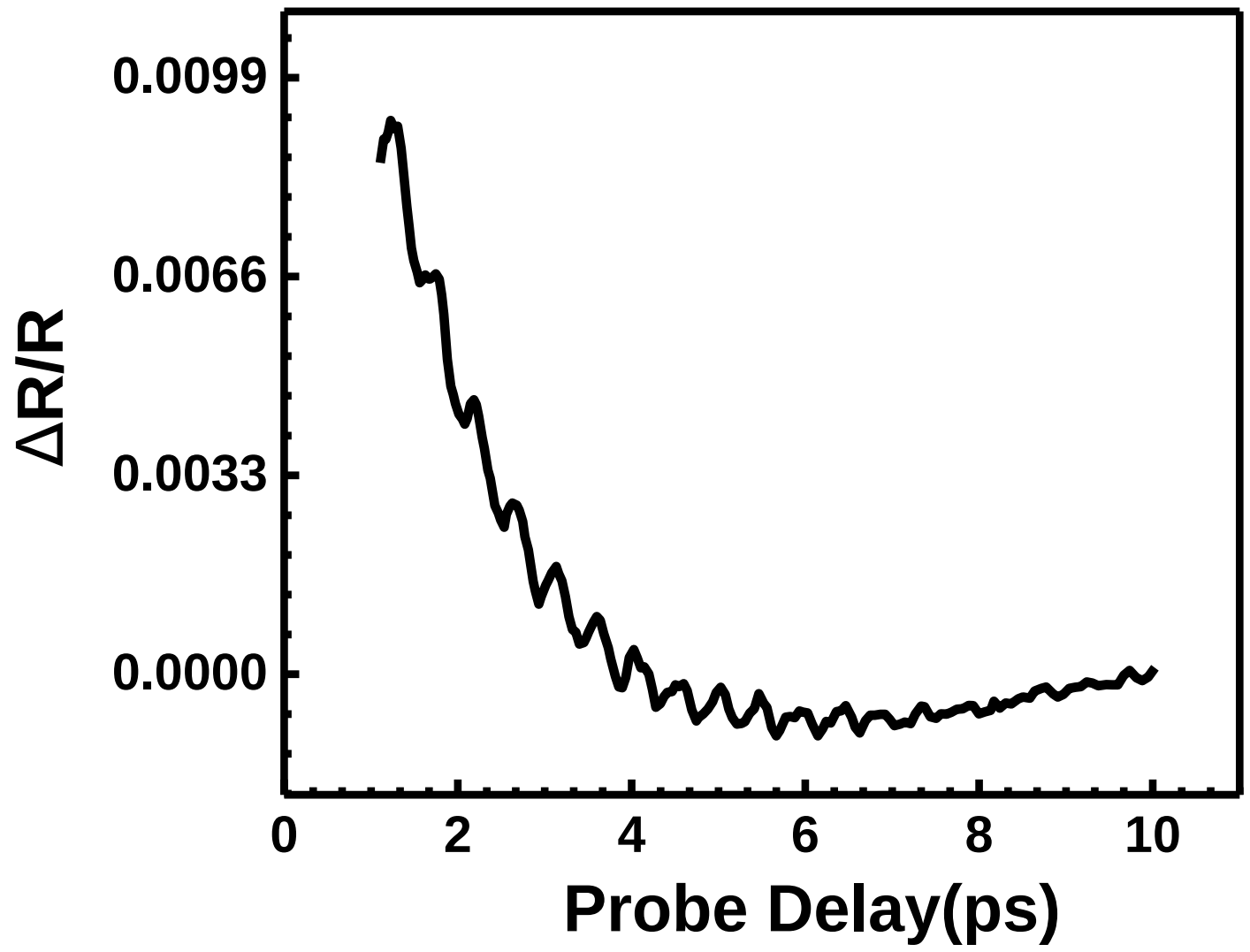

**Figure 6(a)**

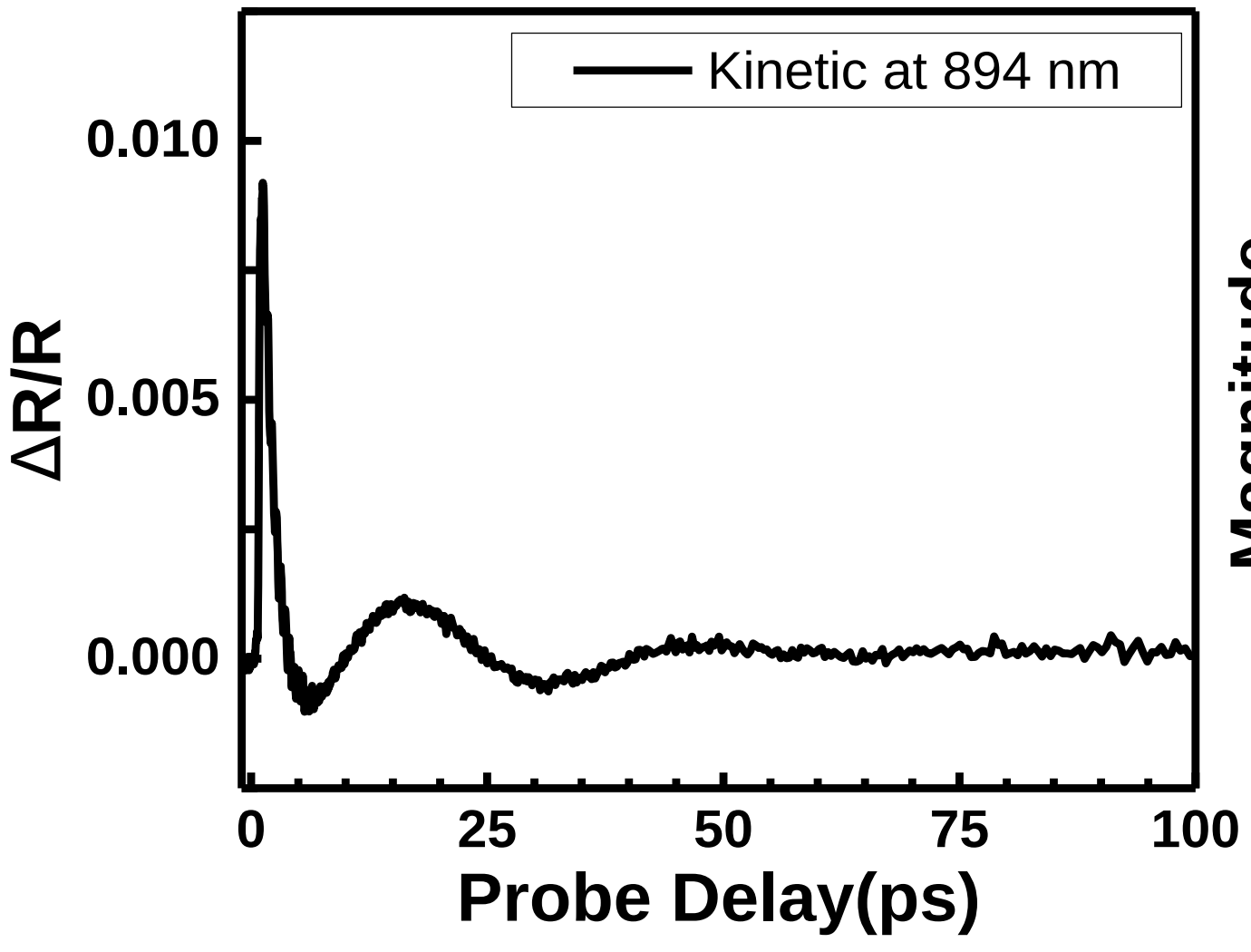

**Figure 6(b)**

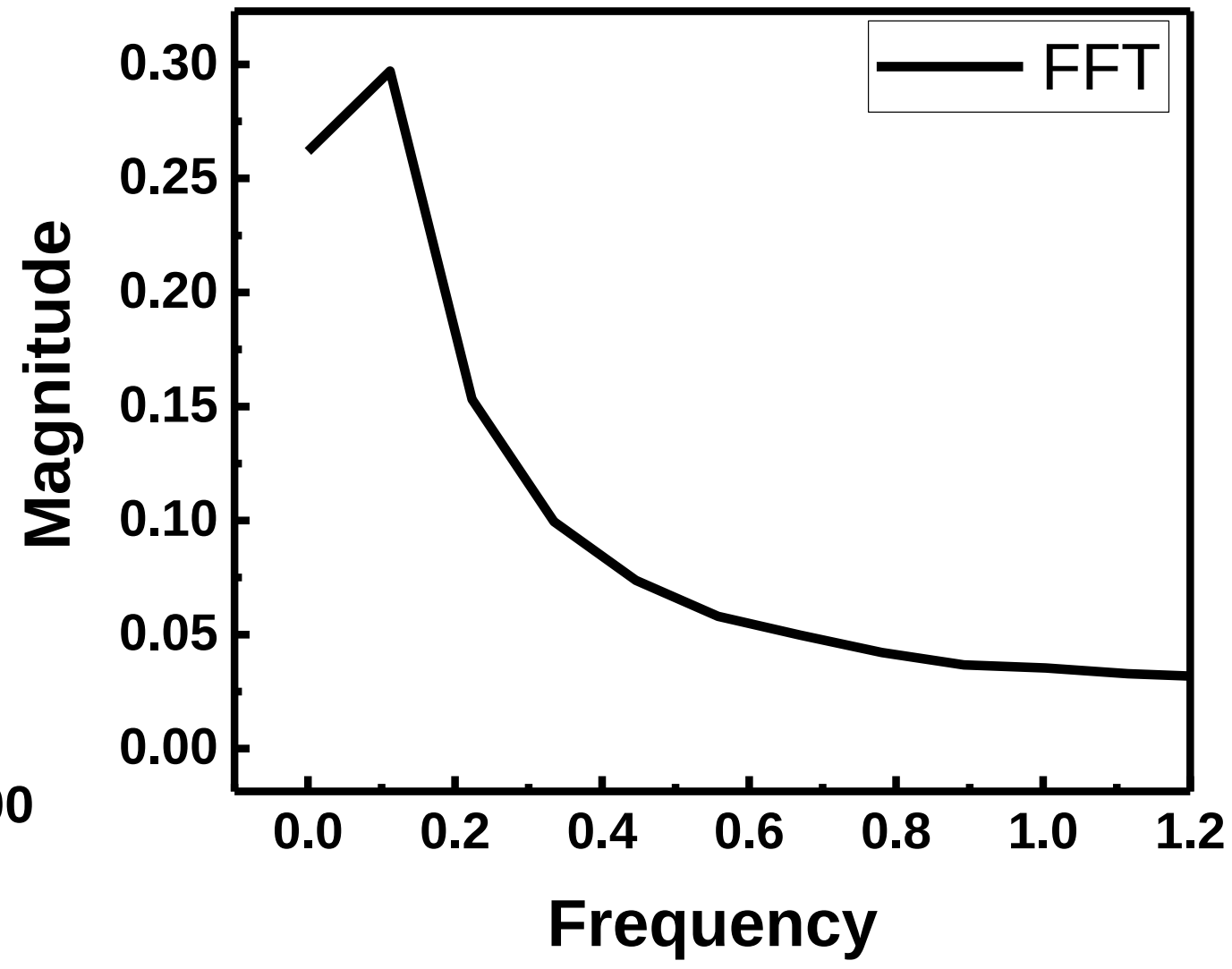

Figure 7(a)

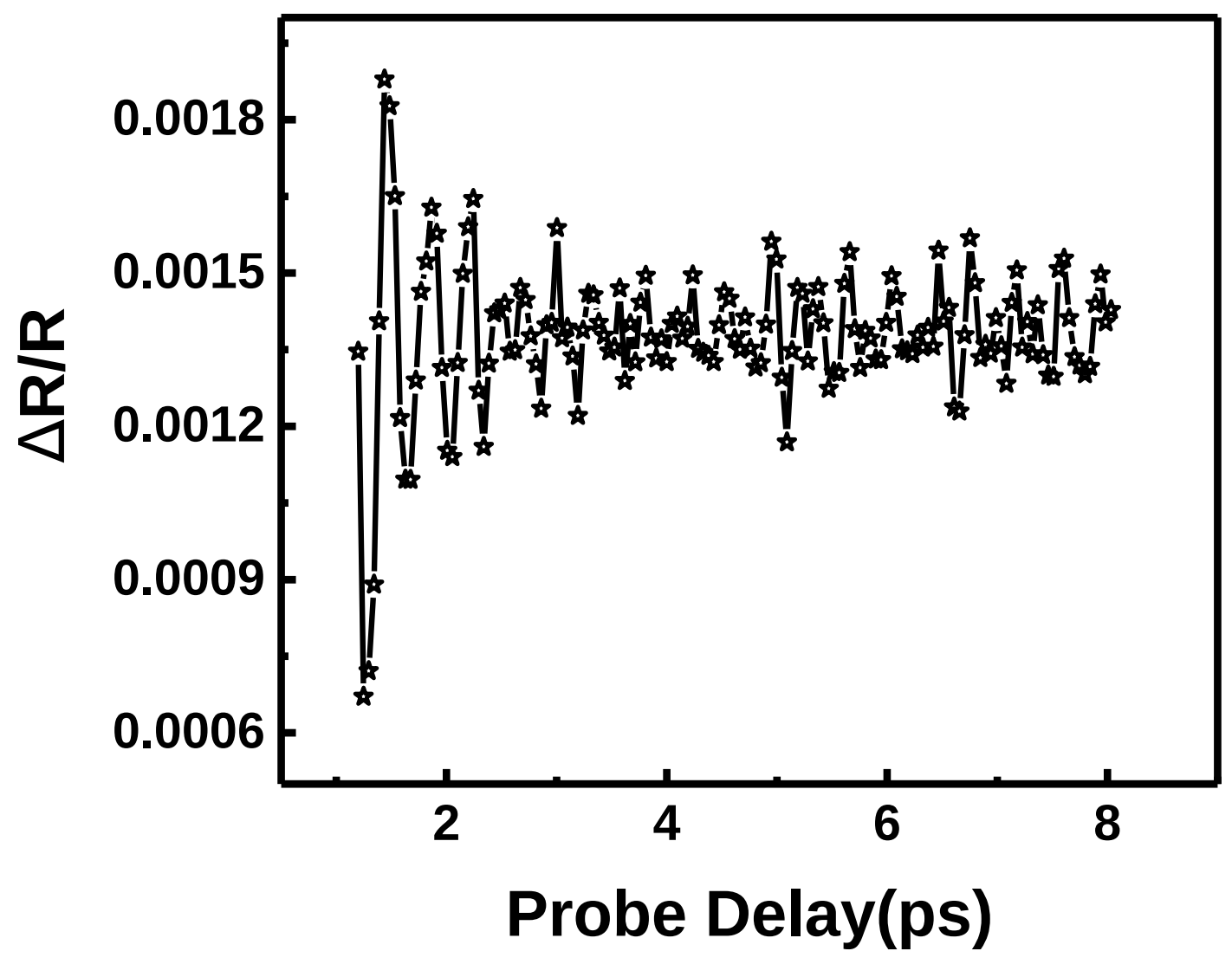

Figure 7(b)

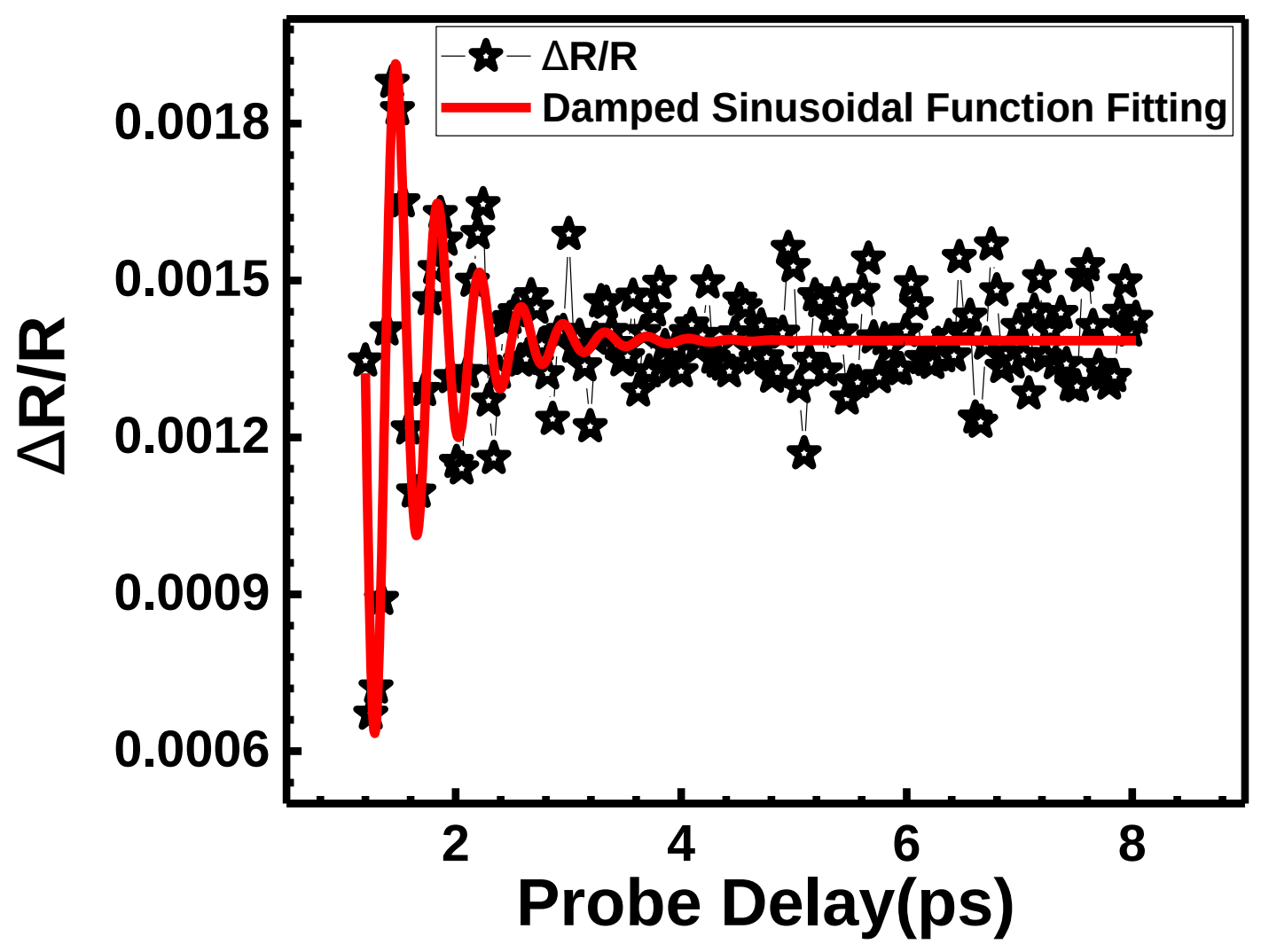

Figure 7(c)

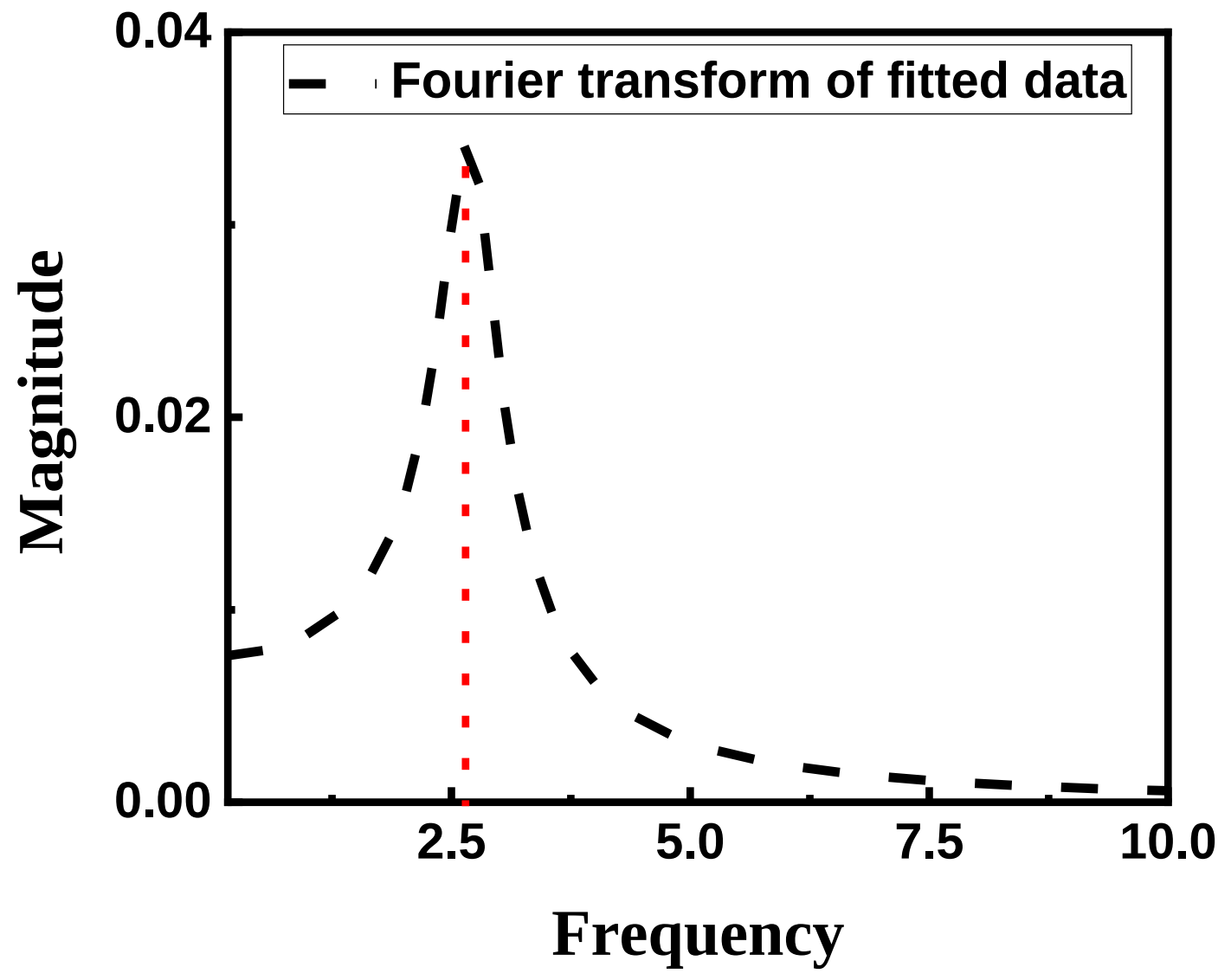

Figure 8

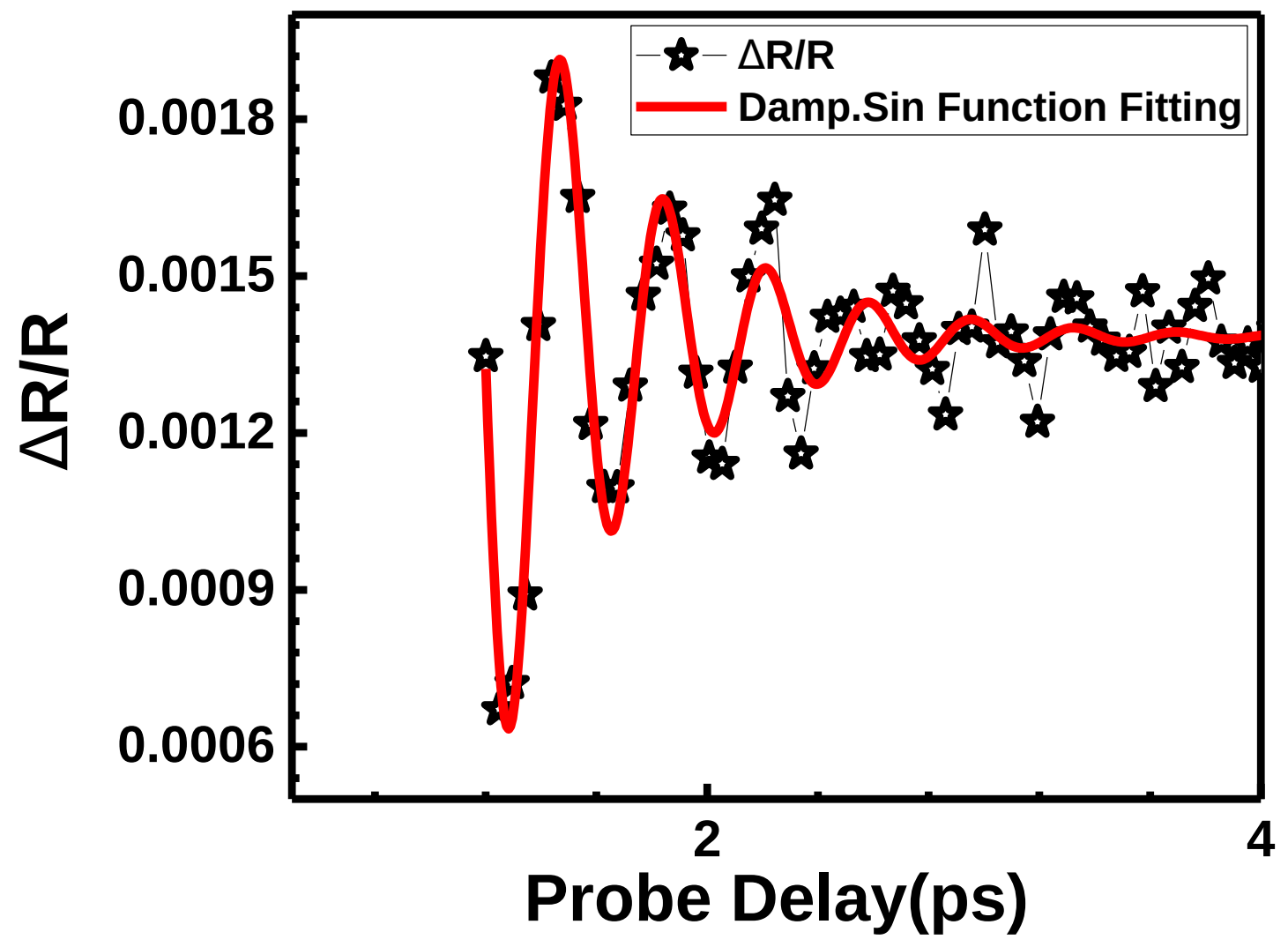

Table 1: Damped Sinusoidal function fitting parameters

| **Model** | Sine-Damp |
|---|---|
| **Equation** | y=y0 + A*$e^{\frac{-x}{t_0}}$*sin(π*(x-xc)/w) |
| **Plot** | D |
| **y0** | 0.00139 ± 6.54715E-6 |
| **xc** | -0.1026 ± 0.01699 |
| **w** | 0.18555 ± 0.00203 |
| **t0** | 0.53041 ± 0.0577 |
| **A** | 0.00848 ± 0.0026 |
| **Reduced Chi-Square** | 6.16586E-9 |
| **R-Square(COD)** | 0.71512 |
| **Adj. R-Square** | 0.70698 |

## REFERENCES

1. Keimer, B., and J. E. Moore. "The physics of quantum materials" Nature Physics **13**, 1045 (2017)
2. Moore, Joel E., and Leon Balents. "Topological invariants of time-reversal-invariant band structures." Physical Review B **75**, 121306 (2007)
3. Kane, Charles L., and Eugene J. Mele. "Z 2 topological order and the quantum spin Hall effect." Physical review letters **95**, 146802 (2005)
4. Hasan, M. Zahid, and Charles L. Kane. "Colloquium: topological insulators." Reviews of modern physics **82**, 3045 (2010)
5. Hsieh, David, Dong Qian, Lewis Wray, YuQi Xia, Yew San Hor, Robert Joseph Cava, and M. Zahid Hasan. "A topological Dirac insulator in a quantum spin Hall phase." Nature **452**, 970 (2008)
6. Qi, Xiao-Liang, Taylor L. Hughes, and Shou-Cheng Zhang. "Topological field theory of time-reversal invariant insulators." Physical Review B **78**, 195424 (2008)
7. Ando, Yoichi. "Topological insulator materials." Journal of the Physical Society of Japan **82**, 102001 (2013)

8. Fu, Liang, and Charles L. Kane. "Topological insulators with inversion symmetry." Physical Review B **76**, 045302 (2007)

9. Shen, Shun-Qing, Wen-Yu Shan, and Hai-Zhou Lu. "Topological insulator and the Dirac equation." In Spin, vol. **1**, 33, 2011

10. Xue, Qi-Kun. "Nanoelectronics: A topological twist for transistors." Nature nanotechnology **6**, 197 (2011)

11. Tokura, Yoshinori, Masashi Kawasaki, and Naoto Nagaosa. "Emergent functions of quantum materials." Nature Physics **13**, 1056-1068 (2017)

12. Yan, Binghai, and Shou-Cheng Zhang. "Topological materials." Reports on Progress in Physics **75**, 096501 (2012)

13. Sobota, Jonathan A., Shuolong Yang, James G. Analytis, Y. L. Chen, Ian R. Fisher, Patrick S. Kirchmann, and Z-X. Shen. "Ultrafast optical excitation of a persistent surface-state population in the topological insulator Bi 2 Se 3." Physical review letters **108**, 117403 (2012)

14. Luo, C. W., H-J. Chen, H. J. Wang, S. A. Ku, K. H. Wu, T. M. Uen, J. Y. Juang et al. "Ultrafast dynamics in topological insulators." In Ultrafast Phenomena and Nanophotonics XVII, vol. **8623**, 86230D, 2013

15. Schnyder, Andreas P., Shinsei Ryu, Akira Furusaki, and Andreas WW Ludwig. "Classification of topological insulators and superconductors in three spatial dimensions." Physical Review B **78**, 195125 (2008).

16. Giannetti, Claudio, Massimo Capone, Daniele Fausti, Michele Fabrizio, Fulvio Parmigiani, and Dragan Mihailovic. "Ultrafast optical spectroscopy of strongly correlated materials and high-temperature superconductors: a non-equilibrium approach." Advances in Physics **65**, 58-238 (2016).

17. Sim, Sangwan, Matthew Brahlek, Nikesh Koirala, Soonyoung Cha, Seongshik Oh, and Hyunyong Choi. "Ultrafast terahertz dynamics of hot Dirac-electron surface scattering in the topological insulator $Bi_2Se_3$." Physical Review B **89**, 165137 (2014)

18. Zhu, Siyuan, Yukiaki Ishida, Kenta Kuroda, Kazuki Sumida, Mao Ye, Jiajia Wang, Hong Pan et al. "Ultrafast electron dynamics at the Dirac node of the topological insulator $Sb_2Te_3$." Scientific reports **5**, 13213 (2015)

19. Mentink, J. H., Karsten Balzer, and Martin Eckstein. "Ultrafast and reversible control of the exchange interaction in Mott insulators." Nature communications **6**, 6708 (2015)

20. Sánchez-Barriga, J., E. Golias, A. Varykhalov, J. Braun, L. V. Yashina, R. Schumann, J. Minár, H. Ebert, O. Kornilov, and O. Rader. "Ultrafast spin-polarization control of Dirac fermions in topological insulators." Physical Review B **93**, 155426 (2016)

21. Kastl, Christoph, Christoph Karnetzky, Helmut Karl, and Alexander W. Holleitner. "Ultrafast helicity control of surface currents in topological insulators with near-unity fidelity." Nature communications **6**, 6617 (2015)

22. Braun, Lukas, Gregor Mussler, Andrzej Hruban, Marcin Konczykowski, Thomas Schumann, Martin Wolf, Markus Münzenberg, Luca Perfetti, and Tobias Kampfrath. "Ultrafast photocurrents at the surface of the three-dimensional topological insulator Bi 2 Se 3." Nature communications **7**, 13259 (2016)

23. Wray, L. Andrew, Su-Yang Xu, Yuqi Xia, Yew San Hor, Dong Qian, Alexei V. Fedorov, Hsin Lin, Arun Bansil, Robert J. Cava, and M. Zahid Hasan. "Observation of topological order in a superconducting doped topological insulator." Nature Physics **6**, 855 (2010)

24. Wang, Y. H., D. Hsieh, E. J. Sie, H. Steinberg, D. R. Gardner, Y. S. Lee, P. Jarillo-Herrero, and N. Gedik. "Measurement of intrinsic Dirac fermion cooling on the surface of the topological insulator $Bi_2Se_3$ using time-resolved and angle-resolved photoemission spectroscopy." Physical Review Letters **109**, 127401 (2012)

25. Kumar, Nardeep, Brian A. Ruzicka, N. P. Butch, P. Syers, K. Kirshenbaum, J. Paglione, and Hui Zhao. "Spatially resolved femtosecond pump-probe study of topological insulator Bi 2 Se 3." Physical Review B **83**, 235306 (2011)

26. Zhang, Xiao, Jing Wang, and Shou-Cheng Zhang. "Topological insulators for high-performance terahertz to infrared applications." Physical review B **82**, 245107 (2010)

27. Di Pietro, P., M. Ortolani, O. Limaj, A. Di Gaspare, V. Giliberti, F. Giorgianni, M. Brahlek et al. "Observation of Dirac plasmons in a topological insulator." Nature nanotechnology **8**, 556 (2013)

28. Jenkins, Gregory S., A. B. Sushkov, D. C. Schmadel, N. P. Butch, P. Syers, J. Paglione, and H. D. Drew. "Terahertz Kerr and reflectivity measurements on the topological insulator Bi 2 Se 3." Physical Review B **82**, 125120 (2010)

29. Shuvaev, A. M., G. V. Astakhov, G. Tkachov, C. Brüne, H. Buhmann, L. W. Molenkamp, and A. Pimenov. "Terahertz quantum Hall effect of Dirac fermions in a topological insulator." Physical Review B **87**, 121104 (2013)

30. Lupi: Terahertz spectroscopy of novel superconductors. Advances in Condensed Matter Physics. **2011**, (2011)

31. Ferguson, Bradley, and Xi-Cheng Zhang. "Materials for terahertz science and technology." Nature materials **1**, 26 (2002)

32. Tonouchi, Masayoshi. "Cutting-edge terahertz technology." Nature photonics **1**, 97 (2007)

33. Siegel, Peter H. "Terahertz technology." IEEE Transactions on microwave theory and techniques **50**, 910-928 (2002)

34. Quarti, C., Marchal, N. and Beljonne, D. Tuning the optoelectronic properties of two-dimensional hybrid perovskite semiconductors with alkyl chain spacers. The journal of physical chemistry letters, **9**(12), 3416-3424 (2018)

35. Kovalenko, M.V., Protesescu, L. and Bodnarchuk, M.I. Properties and potential optoelectronic applications of lead halide perovskite nanocrystals. Science, **358**, 745 (2017)

36. Zhou, Tingwei, Ming Wang, Zhigang Zang, and Liang Fang. "Stable Dynamics Performance and High Efficiency of ABX3-Type Super-Alkali Perovskites First Obtained by Introducing H5O2 Cation." Advanced Energy Materials **9**, 29 (2019)

37. Zhou, Tingwei, Yubo Zhang, Ming Wang, Zhigang Zang, and Xiaosheng Tang. "Tunable electronic structures and high efficiency obtained by introducing superalkali and superhalogen into AMX3-type perovskites." Journal of Power Sources **429**, 120 (2019)

38. Sultana, Rabia, Geet Awana, Banabir Pal, P. K. Maheshwari, Monu Mishra, Govind Gupta, Anurag Gupta, S. Thirupathaiah, and V. P. S. Awana. "Electrical, Thermal and Spectroscopic Characterization of Bulk Bi 2 Se 3 Topological Insulator." Journal of Superconductivity and Novel Magnetism **30**, 2031 (2017)

39. Awana, Geet, Rabia Sultana, P. K. Maheshwari, Reena Goyal, Bhasker Gahtori, Anurag Gupta, and V. P. S. Awana. "Crystal Growth and Magneto-transport of Bi 2 Se 3 Single Crystals." Journal of Superconductivity and Novel Magnetism **30**, 853 (2017)

40. Sharma, P., D. Sharma, N. Vashistha, P. Rani, M. Kumar, S. S. Islam, and V. P. S. Awana. "Ultrafast Spectroscopy of Bi2Se3 Topological Insulator." arXiv preprint arXiv:1910.12439 (2019)

41. Richter, W., and C. R. Becker. "A Raman and far-infrared investigation of phonons in the rhombohedral V2–VI3 compounds Bi2Te3, Bi2Se3, Sb2Te3 and Bi2 (Te1− xSex) 3 (0< x< 1),(Bi1− ySby) 2Te3 (0< y< 1)." physica status solidi (b) **84**, 619(1977)